# Charge order, dielectric response and local structure of La$_{5/3}$Sr$_{1/3}$NiO$_4$ system

M. Filippi*, B. Kundys, S. Agrestini, W. Prellier

Laboratoire CRISMAT, CNRS UMR 6508, ENSICAEN,
6 Boulevard du Marechal Juin, 14050 Caen, France

H. Oyanagi
Photonic Division, National Institute of Industrial Science and Technology, Tsukuba Central 2, 1-1-1 Umezono, Tsukuba, Japan

N.L. Saini
Dipartimento di Fisica, Università di Roma "La Sapienza", P. le Aldo Moro 2, 00185 Roma, Italy

E-mail: mfilippi@few.vu.nl

Charge ordering, dielectric permittivity and local structure of La$_{5/3}$Sr$_{1/3}$NiO$_4$ system have been explored X-ray charge scattering, complex dielectric impedance spectroscopy, and extended X-ray absorption fine structure (EXAFS) measurements, made on the same single crystal sample. The local structure measured by the temperature dependent polarized Ni K-edge EXAFS shows significant distortions in the NiO$_2$ planes. These local distortions could be reasonable cause of high dielectric permittivity of the title system ($\varepsilon \approx 100$ at 5K) with the charge ordering in this system being a ferroelectric-like second order transition.

*Present address: Department of Physics and Astronomy, Condensed Matter Physics, Vrije Universiteit, De Boelelaan 1081, 1081 HV Amsterdam, The Netherlands

# I. INTRODUCTION

Stripe correlations, charge inhomogeneities and their implication on the functional properties of transition metal oxide (TMO) perovskites are the recent key issues in the condensed matter physics, with large impact on the development of new materials for advanced applications[1-4]. Among the TMOs, the $La_{2-x}Sr_xNiO_4$, popularly called nickelates, construct good example of such systems with complex interplay between spin-charge-lattice degrees of freedom[5-16]. The nickelates have been studied intensively, however, a substantial amount of work is dedicated to the $La_{5/3}Sr_{1/3}NiO_4$, merely due to the fact that the system appears with robust charge stripe ordering for this doping, making it a suitable system to address various functions of the TMO perovskites. In addition to the charge ordering, the $La_{5/3}Sr_{1/3}NiO_4$ shows a very high dielectric permittivity ($10^4$) in a wide temperature interval, that drops down at low temperature and reveals a substantial dispersion with frequency[15]. Therefore, $La_{5/3}Sr_{1/3}NiO_4$ is not only a model system to study physics of low temperature orders, but could be also considered a promising material for future applications.

Here we have studied the $La_{5/3}Sr_{1/3}NiO_4$ system for its charge ordering and dielectric properties and made an attempt to find a correlation between these functions and the local atomic structure. The system has been characterized for its charge ordering through measurements of the charge-ordering superstructure as a function of temperature. The same sample has been used to determine the dielectric properties and the local structure using respectively the impedance spectroscopy and the extended X-ray absorption fine structure (EXAFS). We have compared the charge ordering superstructure intensity with the dielectric permittivity, revealing a direct correlation. Moreover, from the local structural study on the title system by the polarized EXAFS, compared with the isostructural $La_{2-x}Sr_xCuO_4$ (LSCO), we have found that the $NiO_2$ planes are largely distorted. The large local distortions indicate broken local symmetry of the title system, that could directly explain the anomalous dielectric properties of this system. In addition, the large local distortions could have strong implication on the charge stripe correlations in the system, and need to be taken into account for any realistic model to address correlating functional properties of the nickelates.

## II. EXPERIMENTAL DETAILS

The single crystalline sample of $La_{5/3}Sr_{1/3}NiO_4$ was grown using the floating zone method[8]. Crystal structure was determined by single crystal diffraction. The lattice parameters were determined to be $a$=b=3.8 Å, $c$=12.8 Å at room temperature. Temperature (accuracy of ±1K) dependent diffraction measurements were performed at the crystallography beam-line of the Elettra synchrotron radiation facility in Trieste. The diffraction data were collected in the reflection geometry, with photon energy of 12.4 KeV (wavelength λ~1Å), using a CCD detector assembly. Thanks to the high brilliance source, it was possible to record a large number of weak superstructures due to charge ordering around the main diffraction peaks. The dielectric impedance measurements were carried out in a PPMS Quantum Design cryostat using an Agilent 4248A RLC bridge as a function temperature and frequencies. Indium was used as electrode material for these measurements.

The Ni K-edge absorption measurements were made in different polarizations at the beamline BL13B of the Photon Factory (Tsukuba), using normal incidence geometry. The synchrotron radiation emitted by a 27-pole wiggler source at the 2.5 GeV Photon Factory storage ring was monochromatized by a double crystal Si(111) monochromator and sagittally focused on the sample. The absorption spectra were recorded by detecting the Ni $K_\alpha$ fluorescence photons over a large solid angle using a 100-element Ge-pixel ar-ray detector (PAD)[18]. Standard procedure was used to extract the EXAFS signal from the absorption spectrum[19] and corrected for the X-ray fluorescence self-absorption[20] before the analysis.

## III. RESULTS AND DISCUSSION

*III.1 Charge Ordering Superstructure*

Fig. 1(a) shows a typical x-ray diffraction image (T=100 K) below the charge ordering temperature, recorded by the CCD detector in a selected reciprocal space (a*-b* plane) of the $La_{5/3}Sr_{1/3}NiO_4$ system. The crystal structure was indexed using a tetragonal unit cell (F4/mmm space group). Besides the Bragg peaks the superstructure spots due to the charge ordering can be seen with the characteristic wave vector for a charge modulation $q_{co}(1,\pm\varepsilon,0)$ with $\varepsilon\sim0.33$[6]. The superstructure spots are not circular, but appear diffused along the direction of the modulation, indicating anisotropic correlation length for the charge ordering in the a-b plane. The charge ordering transition can be quantified by monitoring the integrated intensity of the $(2-\varepsilon, 1)$ superstructure peak (indicated by an arrow in figure1(a)) as a function of the temperature.

The integrated intensity is plotted as a function of temperature in Fig. 1(b). The charge ordering appears below the $T_{CO}\sim220$ K. With decreasing temperature, the peak intensity continues to grow down to $T_L\sim100$ K and remains almost temperature independent below this temperature. The large transition width ($\Delta T \sim 120$ K) indicates that the charge ordering is order-disorder like. In addition, the data suggest that the charge ordering is not a first-order transition as in manganites, but can be better described as a continuous transition[8]. It is worth mentioning that the integrated intensity of the superstructure peak was shown to decrease upon cooling below ~200 K, revealed by an earlier X-ray scattering study[14] on the $La_{5/3}Sr_{1/3}NiO_4$ system, inconsistent with the present data. On the other hand, a very good agreement is found with previous neutron diffraction results [8-12-21]. A non-zero integrated intensity above $T_{CO}$ at T=225 K indicates the presence of very weak scattering at the expected position in the reciprocal space due to presence of dynamic and short-range charge fluctuations.

*III.2 Dielectric properties*

Fig. 2 summarizes the dielectric properties of the $La_{5/3}Sr_{1/3}NiO_4$ single crystal. The current-voltage phase shift (see, e.g. Fig. 2a) reveals that, while the sample response is the one of a pure resistor (i.e., phase shift 0 deg) at low frequency-high temperature, it behaves like a pure capacitor (i.e., phase shift -90deg) at low temperature. The low temperature peak in the phase shift (accompanied by an upturn in the modulus of the impedance, see, e.g., Fig 2b), signature of a relaxation process, shows a large dispersion with frequency. Such a frequency dependent relaxation can arise from different extrinsic contributions, like polarization of the electrodes due to the charge injection [22], or oxygen defects[16], internal barrier layer capacitance[23], or even from a true dipolar relaxation.

The high frequency (1MHz) dielectric permittivity is reported as inset in the Fig. 2, along with the integrated intensity of the charge ordering superstructure of figure 1b. The dielectric permittivity is as high as 14000 at high temperature, and drops down to $10^2$ at low temperature. We can also observe a broad peak like feature in the dielectric constant across the charge ordering transition. This suggests that the dielectric response of the title system is clearly correlated with the charge ordering in the $NiO_2$ plane. Something analogous was also observed in the related charge ordered system $La_{1.5}Sr_{0.5}NiO_4$[24]. We also stress that the low temperature-high frequency value of the dielectric constant, which represents the intrinsic dielectric response of the sample, is rather high ($\varepsilon \sim 100$). The results are in good quantitative and qualitative agreement with Park et al.[15]. As to charge injection at the electrodes, Park et al.[15] have claimed that they could get rid of the electrode polarization by evaporating a thin insulating layer between the sample and the electrodes. Since there is good agreement between our measurements on the bare sample and the results of Park et al., this implies that, either a frequency of 1 MHz is high enough to exclude electrode polarization for this sample, or, the idea of putting an insulator between the sample and the electrode may not be valid to eliminate electrode polarization (as pointed out in ref. 22).

*III.3 Local Structure*

In the ferroelectric systems, structural distortion breaks the average crystallographic symmetry (non-centrosymmetric structures), giving rise to local permanent dipoles. On the other hand, the $La_{5/3}Sr_{1/3}NiO_4$ structure is crystallographically centrosymmetric, however, with a large dielectric permittivity. In such a case, study of local structure becomes crucial to find out the cause of the high dielectric permittivity. Similar approach was applied for other related system, $BaTaO_2N$, exhibiting a dielectric constant as high as 5000, although x-ray, neutron and electron diffraction showing the structure being centrosymmetric[25]. In fact, local structure around the Ta atom, studied by EXAFS, revealed large local distortion, consistent with the high dielectric permittivity of the $BaTaO_2N$ system[26]. It is worth recalling that this giant dielectric response is observed in several TMO[27], whose common feature appears to be a high static permittivity.

In the case of the $La_{5/3}Sr_{1/3}NiO_4$ we have focused on the local structure around the Ni in the $NiO_2$ planes, probed by polarized Ni K-edge EXAFS, due to the fact that the low temperature orderings take place in the $NiO_2$ planes. Figure 3 shows Fourier transforms (FTs) of the Ni K-edge EXAFS oscillations for the $La_{5/3}Sr_{1/3}NiO_4$ at 10 K and 300K, weighted by $k^2$. The correspondent EXAFS oscillations are plotted in the inset, showing the good quality of the measurements. The Fourier transforms provide real space partial atomic distribution around the Ni site in the direction of the X-ray polarization, modulated by the phase shift and scattering amplitude of the backscattering atom, as well as multiple scattering. The Ni near-neighbour distances are 1.9 Å (4 in-plane oxygen atoms), 2.2 Å (2 apex O atoms, which are not seen in this polarization) and 3.2 Å (8 La(Sr) atoms in the rock-salt layers), with the peaks at longer distances being due to multiple scattering of the photoelectron excited at the Ni site.

With polarization parallel to the $NiO_2$ plane, the EXAFS signal due to the in-plane Ni-O bonds is well separated from the signal due to longer bonds and can be easily analyzed separately. The EXAFS amplitude depends on several factors as can be seen from the following general equation for polarized K-edge EXAFS[19]:

$$\chi(k) = \frac{m\pi}{h^2} \sum_i 3N_i \cos^2(\theta_i) \frac{S_o^2}{kR_i^2} f_i(k, R_i) e^{-2R_i/\lambda} e^{-2k^2\sigma_i^2} \sin[2kR_i + \delta_i(k)]$$

Here $N_i$ is the equivalent number of neighboring atoms, at a distance $R_i$, located at an angle ($\theta_i$) with respect to the electric field vector of the polarized synchrotron light. $S_0^2$ is an amplitude correction factor due to photoelectron correlation and (also called passive electrons reduction factor), $f_i(k,R_i)$ is the backscattering amplitude, $\lambda$ is the photoelectron mean free path, and $\sigma_i^2$ represent mean square relative displacements (MSRD) determined by the correlated Debye-Waller factor (DWF) of the photoabsorber-backscatterer pairs (Ni and O). Apart from these, the photoelectron energy origin $E_0$ and the phase shifts $\delta_i$ should be known. The above parameters can be either fixed or allowed to vary when an experimental EXAFS spectrum is parameterized. We have used conventional approach for the Ni-O EXAFS analysis considering a single distance for the Ni-O coordination shell ($N_i = 2$), where the MSRD includes the distortion effects, accounting static and dynamic displacements. The phase and amplitude functions were calculated by feff7[28], following by the modeling using WINXAS code[29].

The maximum number of independent parameters for the least squares fit, given by the Shannon's theorem $N_{ind} \sim (2\Delta k\Delta R)/\pi$ is ~10 for the present case. Only the radial distance $R_i$ and the $\sigma_i^2$ were allowed to vary while all other parameters were kept constant (i.e., 2 parameter fit) in the conventional least squares fit to the experimental EXAFS (see e.g., ref. 19). The starting parameters were taken from diffraction experiments. Within experimental uncertainties the average distances R were found to be temperature independent and similar to the one determined by diffraction experiments (R=1.91 Å).

The temperature dependence of the Ni-O MSRD $\sigma^2_{Ni-O}$ is plotted in figure 4. Before commenting, It is useful to recall that the EXAFS measures the broadening of the distance between two atomic sites, i.e. represents the distance broadening between the absorber (Ni) and the scatterer (O). In the presence of static structural disorder, the MSRD could be represented by a temperature independent (static $\sigma_s^2$) and temperature dependent (dynamic $\sigma_d^2$) terms, i.e.,

$$\sigma^2 = \sigma_s^2 + \sigma_d^2$$

The temperature-dependent part $\sigma_d^2$ could be given by the simple correlated Einstein model as[19]:

$$\sigma_d^2 = \frac{\hbar}{2\omega m_r} \frac{1+e^{-\hbar\omega/k_BT}}{1-e^{-\hbar\omega/k_BT}}$$

where $\omega$ is the Einstein frequency for the Ni-O bond, $m_r$ is the reduced mass of the Ni-O pair, and $\Theta_E = \hbar\omega/k_B$ is the Einstein temperature. To have a better idea on the local disorder in the nickelate, we have also plotted the Cu-O MSRD $\sigma^2_{Cu-O}$ for a single crystal of LaCuO$_{4+\delta}$, measured using the same experimental setup and synchrotron run and analyzed with the same procedure (also the k- and R-ranges were kept similar) in Fig. 4. For further comparison, we have also reproduced data for La$_{1.8}$Sr$_{0.2}$CuO$_4$ taken from ref. 30.

The temperature dependence of both the measured MSRD can be fitted (dashed line) with almost similar Einstein frequency for both the materials ($\omega_{Ni-O}$=383 cm$^{-1}$ and $\omega_{Cu-O}$=402 cm$^{-1}$). The striking difference is a large *temperature* offset, $\sigma^2_s$=0.0025 Å$^2$ between the two datasets. It has been already shown that for the Cu-O pair in the La$_{2-x}$Sr$_x$CuO$_4$ (LSCO) the static component is almost negligible[30]. The large value of $\sigma^2_s$ in the nickelate is a mere indicator of large local disorder around the Ni atom. The disorder could be characterized by a complex atomic displacement configuration with a short range order since the diffraction data do not show such distortions. It is also possible that the disorder occurs instantaneously and the EXAFS, being a local and fast (time scale ~$10^{-15}$ s) probe of the atomic pair distribution function is able to see it well. While the nature of the displacements pattern remains an open question, requiring more experimental and theoretical efforts, the available experimental results clearly reveal large local disorder around the Ni atom within the plane. Such a large disorder could have important implication on the functional properties as the charge order and the dielectric response of the nickelates. And, it is reasonable to think that the high dielectric permittivity ($\varepsilon \approx 100$) of this system being due to the large local atomic configurational disorder in the $NiO_2$ plane. Although, it is difficult to assume that the observed dielectric relaxation is not merely due to the Maxwell-Wagner effect induced by electrodes polarization, the present results do not rule out possibility of a true dipolar relaxation. Indeed, a recent paper[31] have shown remarkable dielectric properties of charge ordered nickelates over a large frequency range with relaxation at high frequency. In addition, a recent paper[32] has reported a electronic structure calculations on the title system, concluding that, in order to reproduce the insulating state of the nickelate a large distortion of the $NiO_6$ octahedra should be taken into account. Thus, the present results have direct implication on the electronic structure of this system, providing important experimental feedback.

## IV. SUMMARY

In summary we have studied $La_{5/3}Sr_{1/3}NiO_4$ system by x-ray scattering, dielectric impedance spectroscopy and EXAFS using the same sample with the aim to find a possible correlation between the functional properties and the atomic structure of this system showing potentially important charge order and high dielectric permittivity. The system undergoes a broad charge ordering transition, similar to the order to disorder like. The anomalous dielectric response, characterized by a large value of the low-temperature high-frequency dielectric constant ($10^2$), appears to be related with the charge ordering in the $NiO_2$ plane. The local structure around the nickel atom, studied by resolution polarized Ni K-edge EXAFS reveals large static disorder in the $NiO_2$ plane, suggesting broken local atomic configurational symmetry, that could explain the dielectric properties of this system ($\varepsilon \approx 100$ at low temperature) and the system can be regarded as a polar glass. Also, the charge ordered state could be described by ordered local dipole field with the charge order transition in these materials being characterized by ferroelectric-like order-disorder transition, consistent with glassy nature of the stripe correlations. Thus, the present work appears to provide useful information on the local structure with correlating function of the title system (in particular, the dielectric properties), and could be helpful in designing materials with controlled functions through manipulation of the local disorder.

**Acknowledgments**

This work was carried out in the frame of the STREP CoMePhS (NMP4 CT-2005-517039) supported by the European Community and the CNRS.

**Figure captions**

Figure 1. 2D image of the x-ray diffraction pattern of the $La_{5/3}Sr_{1/3}NiO_4$ crystal, recorded at T=100 K by a CCD detector in a selected region of a*-b* plane (a). The diffraction spots were indexed using a tetragonal unit cell (F4/mmm space group) (a=b≈3.8 Å, c≈12.8 Å). Temperature evolution of the charge order superstructure (2-ε,1) peak intensity (indicated by an arrow) is shown in panel (b), revealing order-disorder like broad transition.

Figure 2. Current-Voltage phase-shift (a) and Modulus of complex dielectric impedance (b) of the $La_{5/3}Sr_{1/3}NiO_4$ sample, measured as a function of temperature at different frequencies. The dielectric permittivity at 1MHz are shown as an inset (panel b), along with the integrated intensity of the charge ordering superstructure (already shown in figure 1b).

Figure 3. Fourier transforms (FTs) of the EXAFS oscillations (shown as inset) measured on the $La_{5/3}Sr_{1/3}NiO_4$ system at 10K and 300K, showing in-plane atomic distribution around the Ni atom. The FTs are performed between $k_{min}$=3 Å$^{-1}$ and $k_{max}$=17Å$^{-1}$ using a Gaussian window and not corrected for the phase shifts, representing raw experimental data. The peaks 1.6 Å are due to in-plane Ni-O bonds, while the peak at 3.2 Å represents Ni-La(Sr) scatterings. The peak at 3.4 Å is due to the Ni-O-Ni multiple scattering. The first shell fit is also shown (dashed line)

Figure 4. Temperature dependence of mean square relative displacements (MSRD) of the Ni-O pair determined by correlated Debye-Waller factors ($\sigma^2_{Ni-O}$), compared with the MSRD of Cu-O in $LaCuO_{4+\delta}$. The dashed line is a fit and represents correlated Einstein model behavior with a temperature independent (static) contribution. Data for $La_{1.8}Sr_{0.2}CuO_4$ (ref. 30) are also shown.

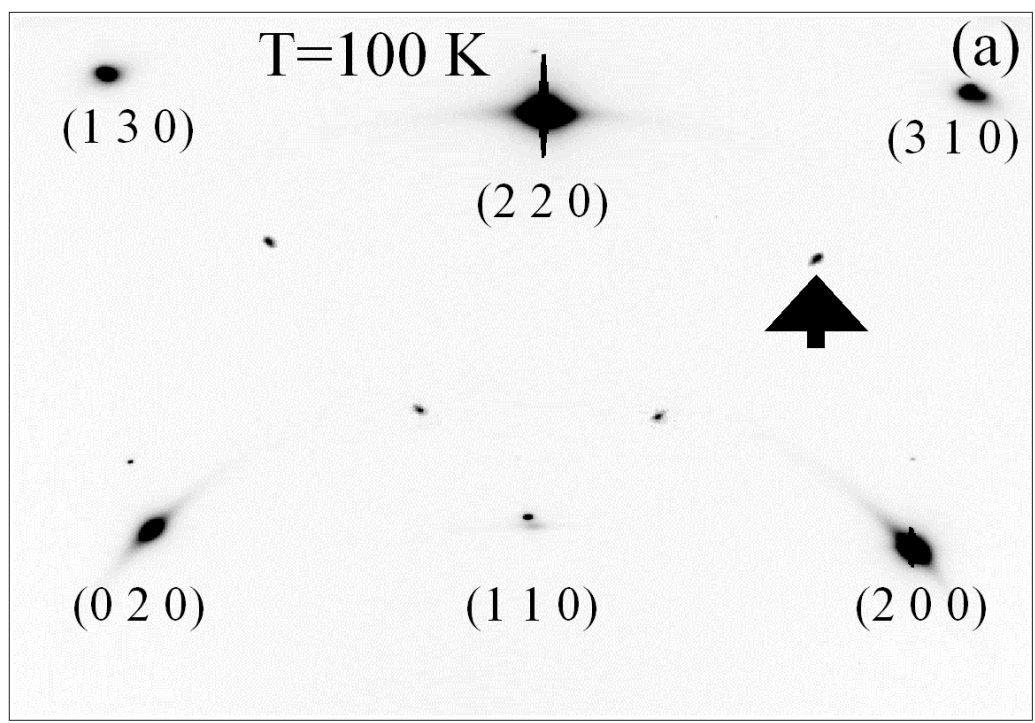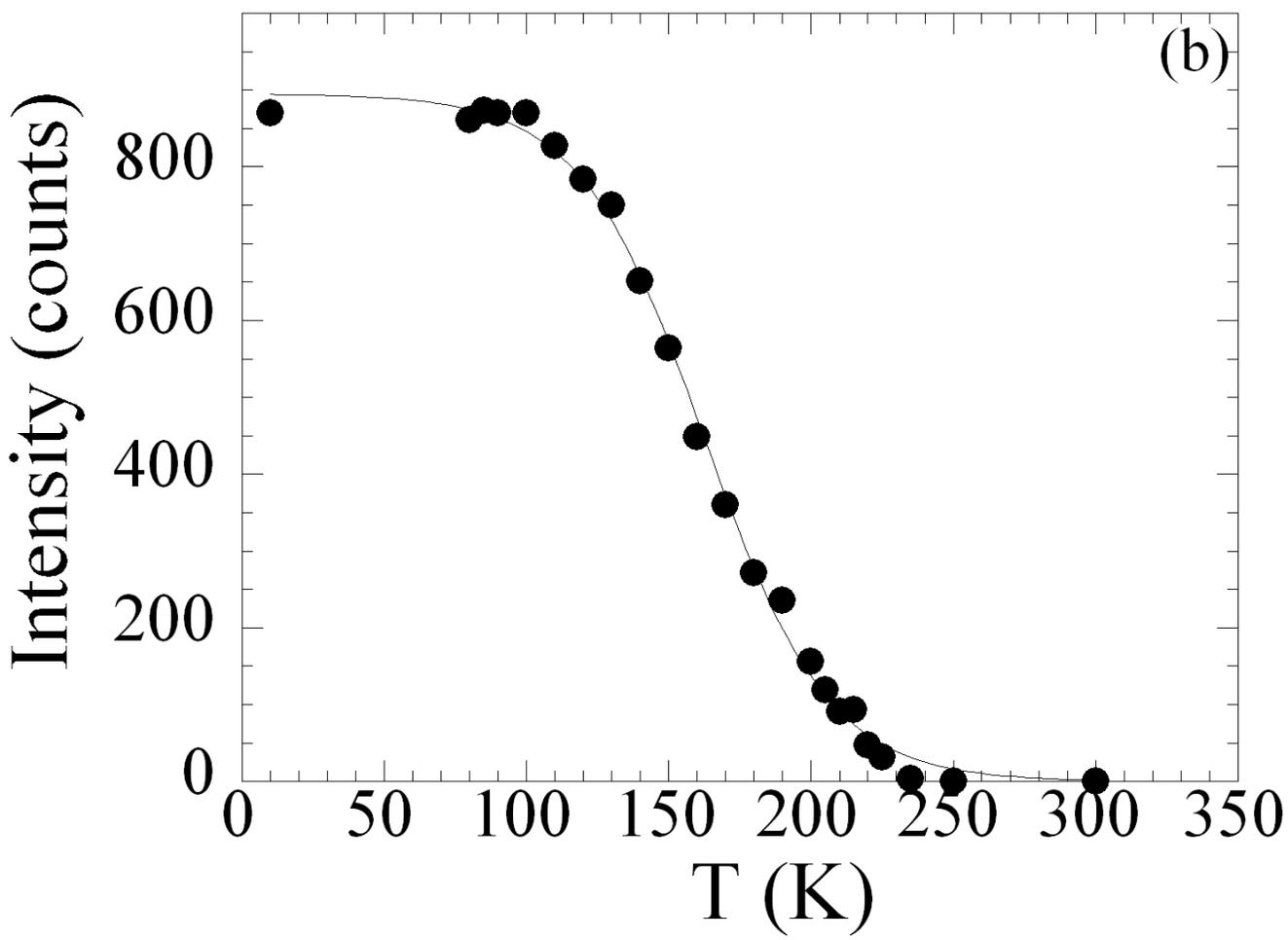

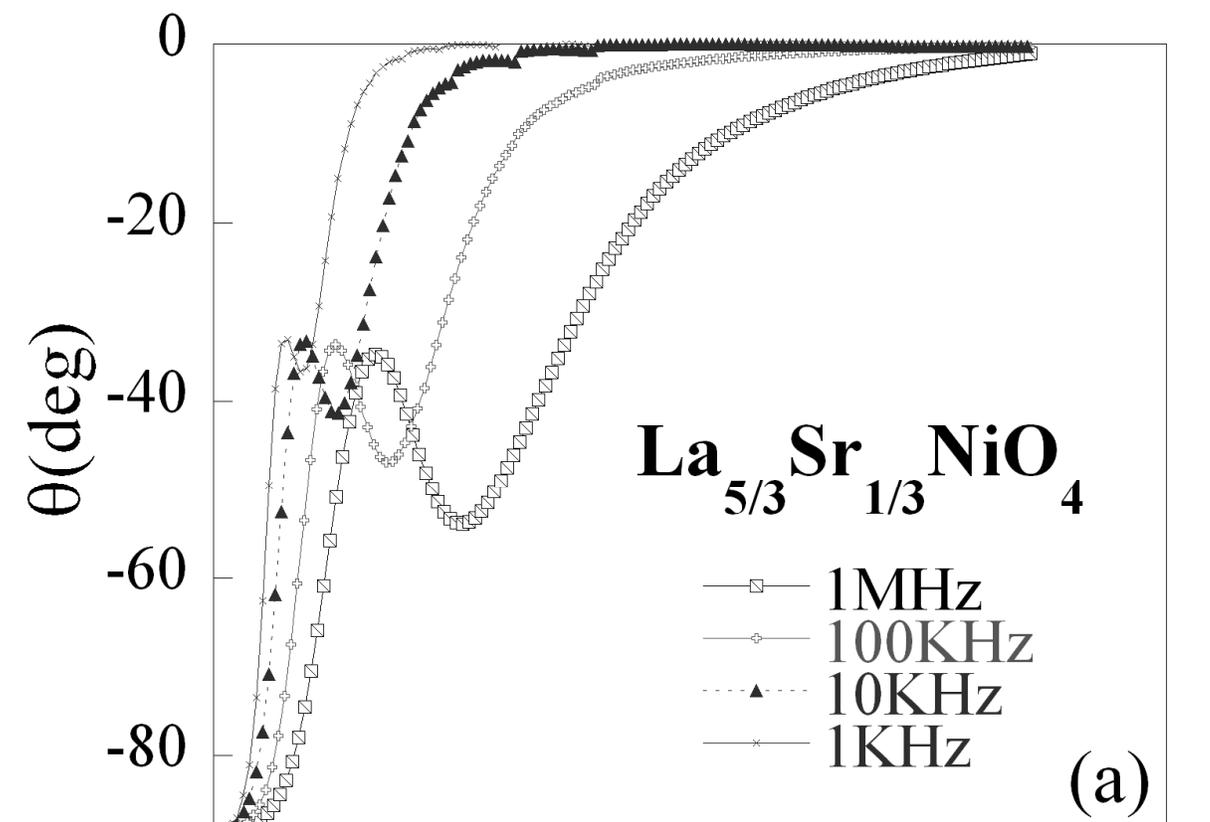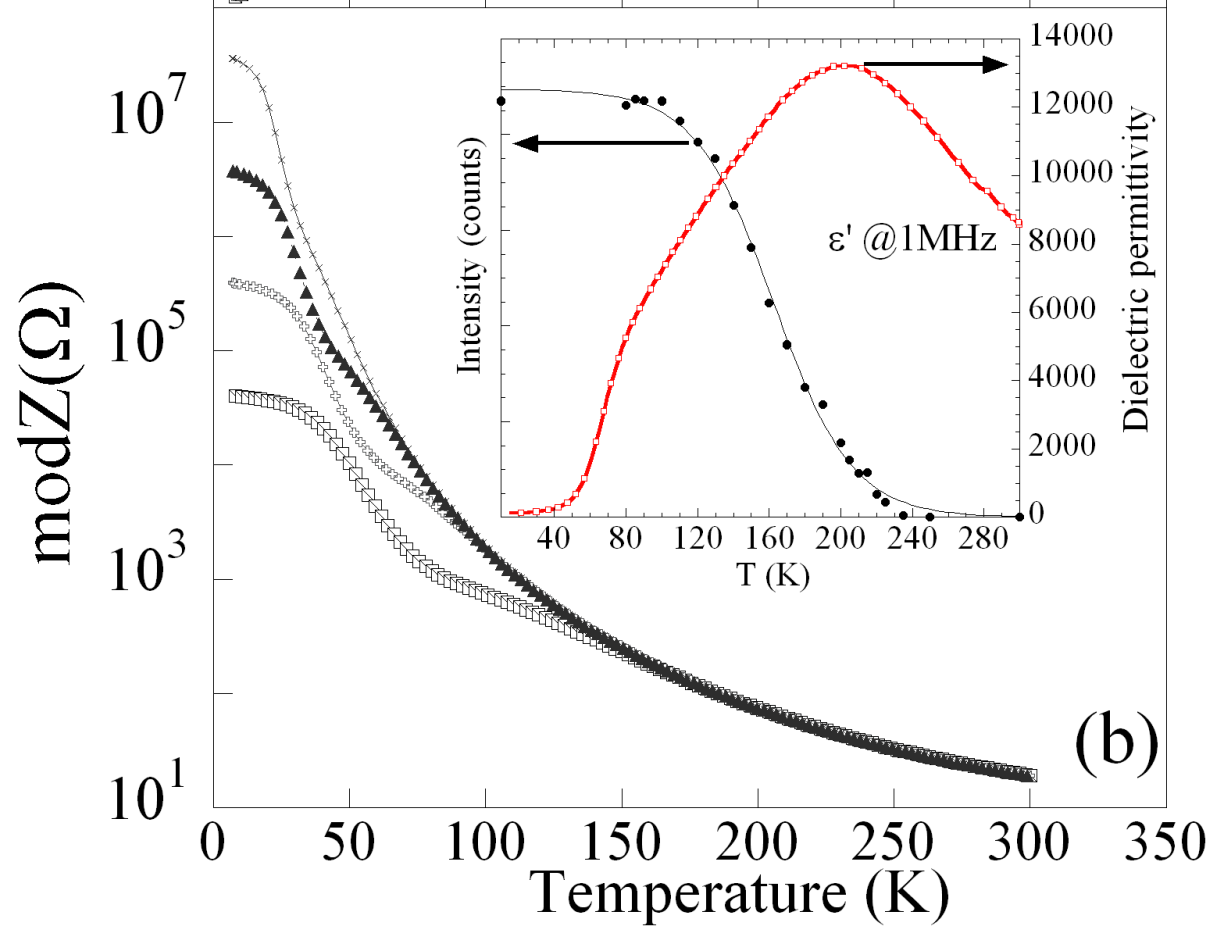

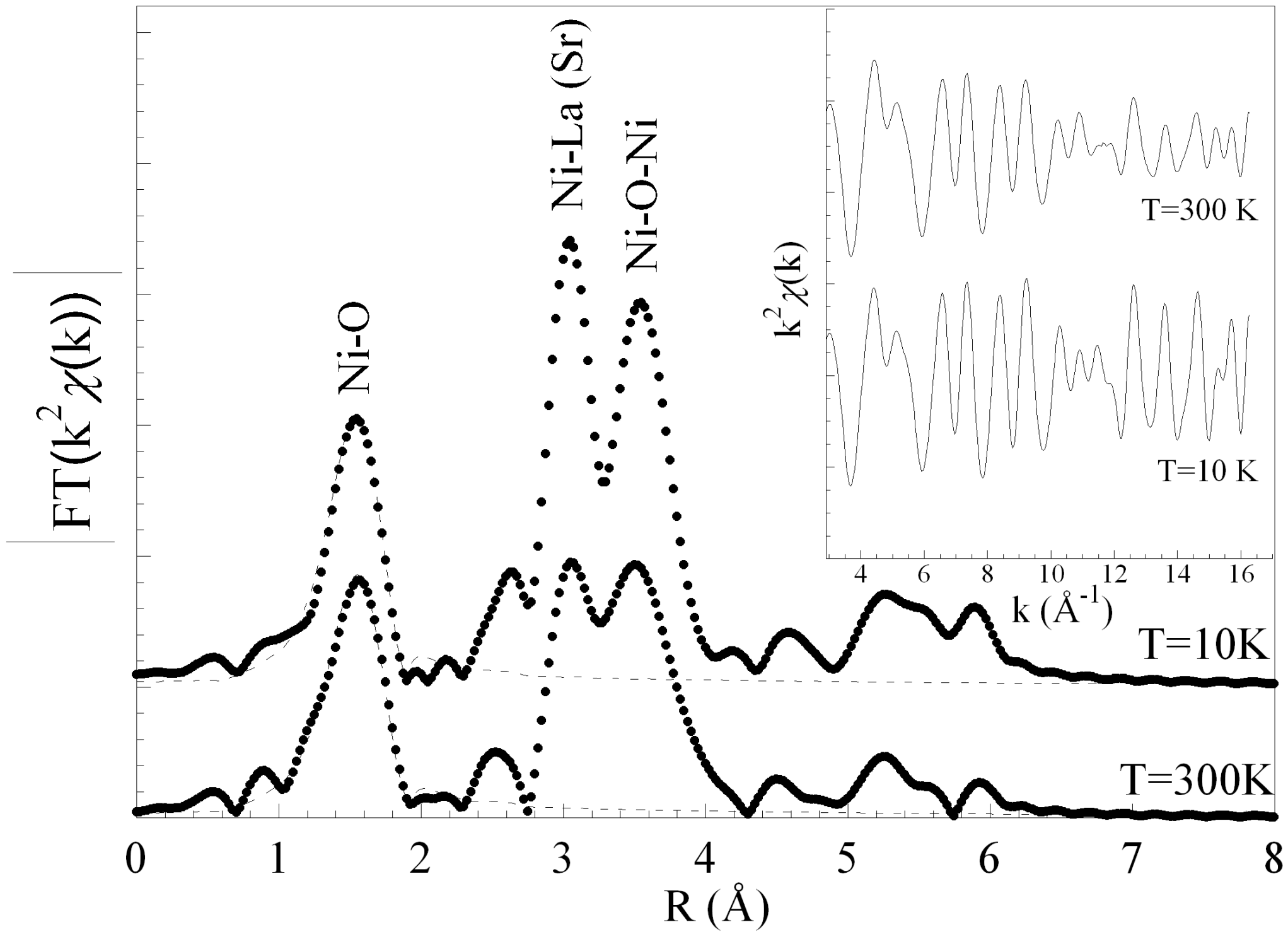

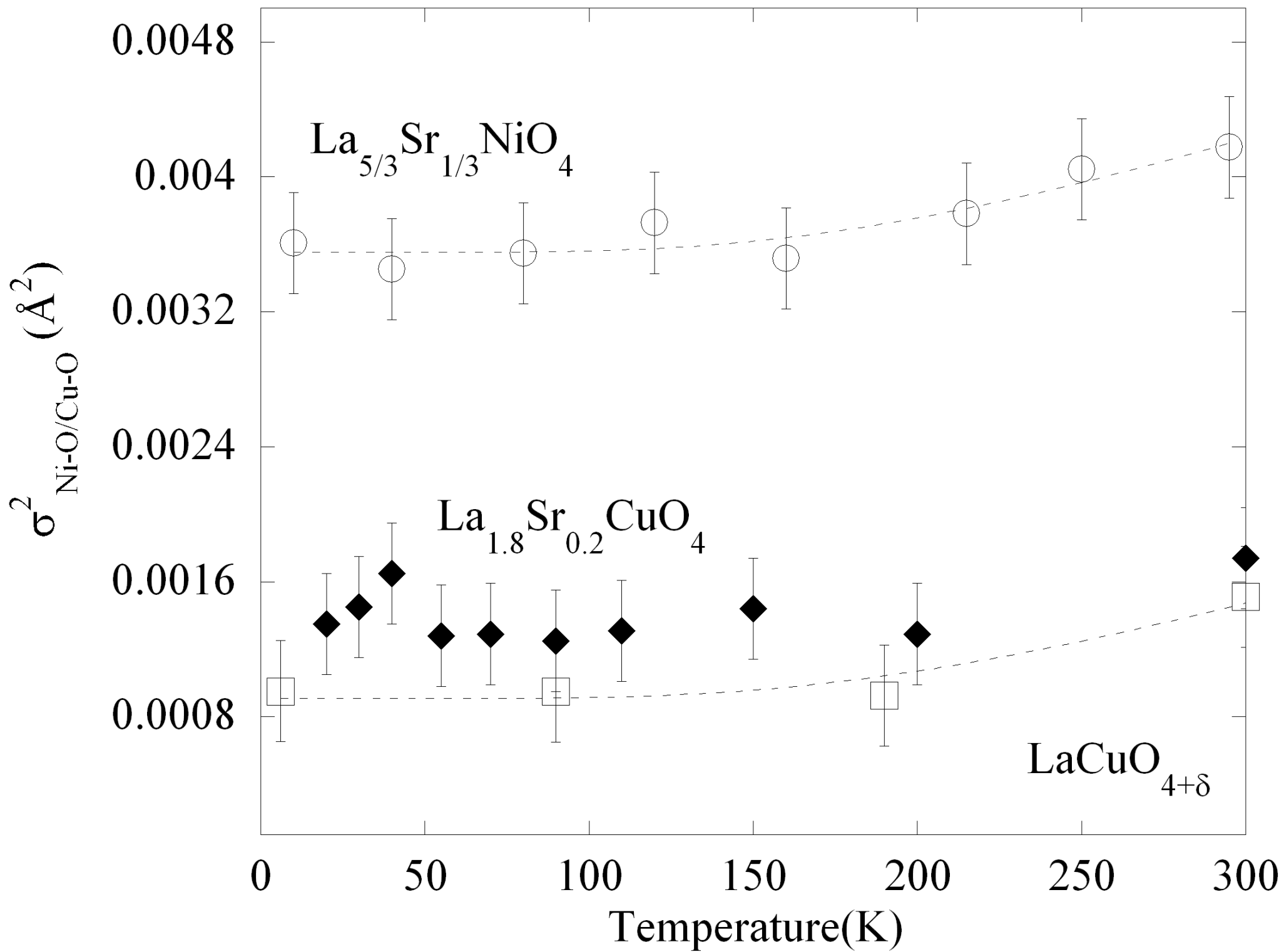